# Size-dependent magnetic properties of Nickel nanochains

Wangzhi Zheng, Lin He, Chinping Chen[a]

Department of Physics, Peking University, Beijing, 100871, PR China

Wei Zhou, Chenmin Liu, Lin Guo[b], Huibin Xu

School of Materials Science and Engineering, Beijing University of Aeronautics and Astronautics, Beijing, 100083, PR China

ABSTRACT

Magnetic properties with 3 different sizes of Ni nanochains, synthesized by a technique of wet chemical solution, have been investigated experimentally. The sample sizes (average diameter of the nano-particles) are 50 nm, 75 nm, and 150 nm with a typical length of a few microns. The characterizations by XRD and TEM reveal that the samples consist of Ni nano-particles forming one dimensional (1D) chain-like structure. The magnetic properties have been investigated by the field-cooled (FC), zero-field-cooled (ZFC) temperature dependent magnetization measurements and the field dependent magnetization (*M-H*) measurement. The results can be well explained within the context of core-shell model. First of all, the freezing of disordered spins in the shell layer have resulted in a peak-like structure on the ZFC curve. The peak position occurs around $T_F \sim$ 13 K. With the 50 nm sample, the field dependent behavior of $T_F(H)$ has been investigated in detail. It is well described by the de Almeida–Thouless (AT) equation for the surface spin glass state. Secondly, the shape anisotropy of 1D structure has caused a wide separation between the FC and ZFC curves. This is mainly attributed to the blocking of the core magnetism. Thirdly, by the *M-H* measurement in the low field region, the open hysteresis loop measured at 5 K < $T_F$ is significantly enlarged in comparison with that taken at $T > T_F$. This indicates that a significant part of the contribution to the magnetic irreversibility at $T < T_F$ is coming from the disordered spins in the shell layer. Lastly, with the reducing sample size, the coercivity, $H_C$, increases, whereas the saturation magnetization goes down dramatically. These imply that, as the sample size reduces, the effect of shape anisotropy becomes larger in the magnetization reversal process and the contribution to the magnetism from the ferromagnetically ordered core becomes smaller.

INTRODUCTION

In recent years, nano-scaled magnetic materials have received much attention, due to the fundamental interest in their unique magnetic properties in comparison with those of the bulk phase and to their promising application potential in technology such as in the high-density magnetic storage, etc. Their magnetic properties are usually dictated by the size, dimension, shape, structure, and morphology of the constituent phases, along with the type and strength of magnetic coupling. Various magnetic nano-structures have been synthesized using different techniques, for example, Co/Ni nanorods via chemical solution [1] or physical vapor-phase epitaxy [2], hollow Ni nanospheres with diameter of 300 to 450 nm via chemical method [3], Ni growing on peptide nanotubes [4], Ni complex nanotubes having diameter of 20 to 40 nm and length of 2 μm [5], and $Co_3O_4$ porous nanotubes with diameter ranging from tens to 200 nm and sidewall thickness ranging from 2 to 20 nm [6]. In virtue of these novel ferromagnetic nanostructures, people increasingly have deeper understanding about the magnetism and their applications. Generally, intensive investigation has been made to demonstrate the enhanced anisotropy on the nano-scaled magnetic material, especially in one or two dimension [7]. Also, the properties receive focused attention include spin disorder in the surface (or boundary) layer [8-10], the magnetic dynamics of single-domain nanoparticles [11], the freezing of spin glass in metal-insulator multilayer [12] and the magnetization reversal mechanism by theoretical [13-16] and experimental investigations [17].

Some experiments have reported that the saturation magnetic moment reduces with decreasing size [8,18,19], while others, obtained the same value as in the bulk phase [20]. By contrast, experiments performed on the Fe, Co, and Ni clusters in a molecular beam have demonstrated that the magnetic moment is atom-like with the cluster size less than 30 atoms. It decreases, approaching the bulk limit as the particle size increases to 700 atoms or so [21]. These seemingly conflicting results indicate that the saturation magnetization of nano-particles depends not only on the magnitude of individual atom or spin moment, but also on the particle size or the complicated surface condition of the particle.

The core-shell model has been applied to investigate the spin glass property in the surface layer of a nanoparticle with single crystal phase. For example, the surface spin glass state with $NiFe_2O_4$ [8], $Fe_3O_4$ [22], or $MnFe_2O_4$ nanoparticles [23] has been studied by applying this model. By this model, the magnetic structure is divided into two major components. The core refers to the inner part of a nanoparticle that exhibits bulk-like magnetic ordering. On the other hand, the magnetic moments in the surface layer often form the spin-glass state, which shows very different magnetic property from that of bulk phase. This is because atoms on the surface are often less coordinated and lack of symmetry. This may consequently complicate the interaction between atoms and form the surface random potential. In addition, as the particle size reduces, the magnetic property of the surface layer is expected to become increasingly important due to the growing surface-to-volume ratio.

With the one dimensional (1D) structure or shape, such as those of nanorods [2,24], or nanowires [25-26], the investigations are often concentrated on the magnetization reversal mechanism and the effect of shape anisotropy. The magnetization reversal associated with the 1D magnetic nano-chain has been investigated theoretically by the model of "chain of spheres" proposed in the early days by Jacobs and Bean [13] and studied in detail by Aharoni[14-16]. Within the framework of this model, only the pair-wise, long-range dipolar interaction among spheres of single magnetic domain is considered. Two of the most important modes, the fanning (or buckling) [13] and the curling[14-16], form a complete spectrum of the magnetization eigenstate. Afterwards, L. Zhang and A. Manthirama have verified the fanning mode experimentally with the ferromagnetic Fe nanochains [17].

Very recently, pure Ni nanospheres connected with each other forming chain structures have been synthesized by a technique of wet chemical route [27]. By controlling the reaction conditions, we can obtain the constituent nanospheres with different diameters, ranging from 50 nm to 150 nm. This provides a unique model system for the fundamental investigation into the magnetism associated with the surface spin, the relation between magnetism and sample size, and the magnetization reversal mechanism in the 1D nanostructure. The thermal behavior of the magnetization, by the FC and ZFC measurements, has exhibited two major features. One is related to the freezing of spins in the surface layer at low temperature, which can be identified by a peak at the freezing temperature, $T_F$, on the ZFC curve. The other is the blocking property of the magnetization in the anisotropic potential attributed to the 1D shape anisotropy of the chain-like structure. The corresponding feature is the blocking temperature, $T_B$, observed with the ZFC curve. The field dependent measurements (*M-H*) at different temperatures show that the saturation magnetization reduces with the reducing particle size. The coercivity of 50 nm nickel nanochain has been determined experimentally at various temperatures. The result can be better described by the theoretical calculation of fanning mode based on the "chain of spheres" model.

## SAMPLE PREPARATION AND CHARACTERIZATION

Synthesis of the sample has been made by the technique of a wet chemical route, the same as that in our previous report [28]. The chain-like structure is formed by a self-assembled process of Ni nanospheres with the modification of a multidentate ligand poly, vinyl pyrrolidone (PVP), on the surface. The PVP is known to have no magnetism. Three different sizes of nano-chain are obtained, including 50 nm, 75 nm, and 150 nm, labeled as S50, S75 and S150, respectively. The size distribution is about 10 nm. The samples have a typical length of a few micrometers. Typical TEM images of the S50, S75 and S150 samples are shown in Fig. 1. In Fig. 1a, the residual PVP is vaguely visible surrounding the Ni chain. The Ni nano-particles are structurally connected together, forming chain-like networks. In particular, the chains in Fig. 1c have almost grown into wires.

Figure 2 shows XRD patterns of the samples S50, S75, and S150. The main peaks, which are indexed with the S150 curve representative for all of the three samples, correspond to the face centered cubic (fcc) nickel (JCPDS 04-0850). The remaining weak peaks marked with stars on the S50 curve conform to the crystal planes of nickel carbide, $Ni_3C$, (JCPDS77-0194). The surface carbonization of the nickel nano-chains occurred during the preparation process, which leads to $Ni_3C$ probed in the XRD measurement. The peaks corresponding to $Ni_3C$ in S50 are a little stronger than the ones in S75. This further proves that $Ni_3C$ indeed exists in the surface. As S50 generally has a larger surface to volume ratio than that of S75, the ratio of the carbonized atoms to the uncarbonized ones in S50 is larger than that in S75. Comparing with the XRD pattern of S50 or S75, the peaks appearing in the S150 curve can be easily indexed as pure fcc nickel (JCPDS 04-0850). No peak corresponding to $Ni_3C$ in S150 has been observed. Perhaps, it is because that S150 has the smallest surface to volume ratio. The amount of $Ni_3C$ existing with S150 is beyond the sensitivity of XRD measurement.

## MEASUREMENTS AND ANALYSIS

The magnetic measurements on the powdered samples have been performed using SQUID magnetometer (Quantum Design). The temperature dependent magnetizations are recorded in an applied field, $H_{app}$ = 90 Oe, after zero field cooling (ZFC) or field cooling (FC) from 300 K to 5 K. By the FC mode, the cooling field is set as $H_{COOL}$ = 20 kOe. For all the three samples, the two branches of FC and ZFC curves separate from each other with the temperature going up to 395 K, as shown in Fig. 3. It evidences a blocking behavior of the core magnetism. An anisotropic potential barrier blocks the magnetization reversal process. This potential barrier is attributed to the shape anisotropy of the 1D chain-like structure. Even when the temperature reaches 395 K, the thermal activation energy is insufficient for the blocked magnetization to overcome the potential barrier. In the FC measurement, the magnetization initially aligned by the external cooling field at high temperature is "trapped" by the anisotropic barrier as the temperature goes down. During the measurement, the cooling field, ~ 20 kOe, is replaced by a small recording field, 90 Oe. With the ascending temperature, $M_{FC}(T)$ reduces due to the thermal activation effect. On the other hand, with the ZFC, the magnetization of each magnetic core tends to be "quenched" in the spatially random orientation during the cooling process. Hence, the net magnetization is very small. When the temperature increases, $M_{ZFC}$ increases because the thermal activation effect tends to re-align the magnetization of each magnetic core toward the external field direction. The magnitude of $M_{FC}$ shows significant size effect. It becomes smaller with the reducing sample size. Another interesting feature has been observed with the ZFC curves. In the low temperature region, $M_{ZFC}$ exhibits a peak structure at about 13 K, see the inset of Fig. 3. There is no obvious dependence of the peak position on the sample size. The peak height, however, increases with the reducing sample size. This suggests that it is resulting from the thermal activation of the frozen moment in the surface layer. The peak temperature is defined as the freezing temperature, $T_F$.

In order to further investigate the effect of applied field on the blocking and freezing property of the sample, the ZFC curves for S50 have been recorded by the fields of 90 Oe, 200 Oe, 500 Oe, 800 Oe, and 1 kOe. The FC curve measured by 1 kOe has been included as well for the purpose of comparison. These results are plotted in Fig. 4a. As the applied field for the measurement increases, a softening-like downward trend becomes more obvious in the high temperature region. It is owing to the effect of thermal activation on the core magnetism overcoming the blocking potential barrier. The blocking temperature, $T_B$ is defined by the broad maximum of the magnetization in the ZFC curve. As the applied field increases, $T_B$ moves toward the low temperature end. It vanishes by the applied field of 1 kOe. The ZFC and FC curves almost coincide with each other in the applied field of 1 kOe, except in the low temperature region, ~ 10 K, where the ZFC curve still exhibits a peak showing the feature of freezing property. The freezing temperature, $T_F$, shifts down slightly with the progressively increasing applied magnetic field. The competition among the effects of thermal activation, blocking potential and applied field strength is further revealed in the susceptibility plot, $\chi(H) = M_{ZFC}(H)/H$, versus temperature in Fig. 4b. At $T < 260$ K, the curves recorded by 90 Oe and 200 Oe almost coincide with each other, going up linearly with the temperature. However, the data recorded by 200 Oe has shown a softening feature at high temperature. It departs from the linearity at 260 K. This indicates that, for the magnetization initially blocked by the anisotropic potential, the thermal activation energy has become comparable to the anisotropic potential energy under the effect of applied field. When the field increases to 500 Oe, the temperature of deviation from the linearity drops to about 115 K. As the field increases further to 800 Oe and 1 kOe, the effect of magnetic field would dominate the anisotropic energy and the thermal activation effect. It is noted that the sample has been kept at room temperature for a few days before every run so that the magnetism can fully relax to ensure a reliable ZFC measurement.

The detailed investigation into the S50 clearly shows that the surface spins lead to a typical freezing behavior at low temperature. It gives rise to the peak structure around 13 K in Fig. 4a. The relation between the freezing temperature, $T_F$, and the corresponding external applied field, $H = H_{AT}$, can be described by the de Almeida–Thouless (AT) equation, see for example reference [22],

$$H_{AT}(T_F) \propto \left(1 - \frac{T_F}{T_f}\right)^{3/2}. \qquad (1)$$

$T_F$ is the position of the observed low temperature peak, and $T_f$ is a fitting parameter representing the freezing temperature with vanishing magnetic field. The experimental data of the freezing temperature versus the external field are illustrated in Fig. 5a. The error bars represent the uncertainty in determining $T_F$. It is 0.5 K for each point, estimated from the separation of data points. The dashed line represents the fitting result by the AT equation. It goes through the data points, showing a reasonably good fitting. The $H^{2/3}$ dependence demonstrates the "surface" spin-glass behavior, which is different from the $H^{1/2}$ dependence for a "volume" spin-glass behavior as discussed in reference [22]. The zero field freezing temperature is obtained from Fig. 5a by extrapolation as $T_f = 14.5$ K.

Besides the freezing temperature, $T_F$, obtained from Fig. 4a, the blocking temperature, $T_B$, can also be determined from the broad maximum in the high temperature region of the ZFC curves. It is plotted versus the applied field in Fig. 5b. In the case of 90 Oe, the ZFC curve does not reach the maximum within the range of measurement. Hence, $T_B$ should occur between 400 K and the Curie point of Ni, ~ 631 K. It is therefore reasonable to estimate the value as 520 ± 100 K. Furthermore, by extrapolation of the ZFC and FC curves for S50 in Fig. 3, the intersection is expected to fall within this estimated value. On the other hand, the ZFC curve does not show any maximum in the applied field of 1 kOe except for the freezing peak at $T_F$. One can estimate $T_B$ by the criteria that it is close to the temperature where FC and ZFC merge. It is about 50 ± 10 K A simple equation based on the analysis of Neel [29] and Brown [30] for the blocking temperature, see for example reference [31], is applied to fit these points,

$$T_B(H) = \frac{KV}{k_B \ln(\tau_{obs}/\tau_0)} \left(1 - \frac{H}{H_K}\right)^{3/2} \qquad (2)$$

where $H_K$ is a fitting parameter representing the anisotropic field, $K$ is the anisotropic constant, $V$ is the volume of the particles, $\tau_{obs}$ is the experimental observation interval, and $\tau_0$ is the characteristic time constant usually in the range of $10^{-9} \sim 10^{-11}$ sec. The anisotropic field can be expressed as $H_K = 2K/M_S$, $M_S$ is the saturation magnetization. When applying equation (2) for the fitting, we use $\ln(\tau_{obs}/\tau_0) = 25$, $M_S$ = 485 emu/cm$^3$ (bulk value of Ni), and leave $K$ and $V$ as fitting parameters. The fitting result is presented in Fig. 5b by the dashed curve. The anisotropic constant $K$ we obtain from the fitting is about $2.7\times10^6$ erg/cm$^3$, which is two orders in magnitude larger than that of the bulk phase, $4.5\times10^4$ erg/cm$^3$. This is not surprising from the quasi-one-dimensional shape anisotropy. Since even with thin film, $K$ is one to two orders in magnitude larger than that of the corresponding bulk phase [32,33]. Using this fitting value of $K$ and the bulk value of $M_S$, we obtain $H_K \sim 1.1$ kOe. This result is in agreement with the $M(T)$ measurements shown in Fig. 4, in which, the FC and ZFC curves almost coincide with each other under the applied field of 1 kOe. This indicates experimentally that $H_K$ is on the order of 1 kOe. It is noted that a modified equation for $T_B$ and $H$ other than equation (2) is also presented in reference [31] to take into account the interaction within the framework of random anisotropy model (RAM). It turns out that it does not fit our experimental data well. This implies that the inter-nanochain interaction is negligible.

The field dependent magnetization ($M$-$H$) measurements at 5 K, 300 K, and 380 K, on S50 are plotted in Fig. 6. At $T = 5$ K, below the freezing temperature, the loop is much more pronounced in comparison with the ones measured at high temperatures. This indicates an important contribution from the surface magnetism. The saturation magnetization determined in the high field region $\sim 50$ kOe, see the inset, are 25 emu/g (5 K), 11.0 emu/g (300 K), and 10.8 emu/g (380 K), corresponding to the effective moment of 0.263 $\mu_B$, 0.116 $\mu_B$, and 0.113 $\mu_B$ per Ni atom, respectively. This is much smaller than the listed bulk value of 0.606 $\mu_B$ per Ni atom at 300 K. The much enhanced moment obtained at $T < T_F$ in comparison with that at $T > T_F$ suggests that the surface magnetism is indeed very important for the nano-sized particles. This is in consistent with the results in the ZFC and FC measurements. The field dependent measurements have been carried out for S75 and S150 as well. The saturation magnetizations and the coercivity are summarized in Tab. 1. The larger the sample size, the larger the saturation magnetization. For S150 at 300 K, the value is 46 emu/g, reaching 80 % of the bulk limit. This suggests that the magnetism starts deviating from the bulk value at the sample size around 150 nm or so. On the other hand, the coercivity reduces with the increasing sample size. In Fig 7, the coercive field, $H_C$, determined from the loops in Fig. 6 is plotted along with $\Delta M = M_{FC} - M_{ZFC}$ for S50. These two exhibit a proportional relation. It further indicates the important correlation of the surface magnetism with the hysteresis loop.

The magnetization reversal mechanism in the present nanochain structure can be described by the model of "chain of spheres" [13]. With the fanning mode, the $H_C$ is expressed as,

$$H_{C,n} = \frac{\mu}{R^3}(6K_n - 4L_n) \qquad (3)$$

where $\mu$ and $R$ are the dipole moment and the diameter of each sphere. The expression in the parenthesis accounts for the dipolar interaction between each pair of magnetic spheres based on the assumption of fanning mode, and the number, n, is for the number of spheres in the chain [13]. The term $\mu/R^3$ can be expressed, using saturation magnetization per unit volume $m_S$, as $\mu/R^3 = (\pi/6)m_S$. Thus, by using the experimental value of $M_S$, the density of bulk Ni, and assuming n =12, we obtain the values of $H_C$ for S50 as 295 Oe at 5 K, 130 Oe at 300 K, and 112 Oe at 380K. In order to make a comparison, we also estimate the coercivity according to the coherent rotational mode. Using the expression in reference [13], we obtain the values 726 Oe (5 K), 320.Oe (300 K), and 276 Oe (380 K). The result predicted by the coherent rotational mode is much larger than the experimental value listed in Tab. 1, while the fanning mode gives a more reasonable result to depict the magnetization reversal in the nickel nanochain. A point worthy of mentioning is that the calculated coercivity for S50 by the fanning mode is smaller than the experimental value, 392 Oe at 5 K, 139 Oe at 300 K, and 97 Oe at 380 K. In particular, the calculated value at 5 K accounts for only 75 % of the experimental one. This is not surprising since at 5 K, the core radius is smaller than the apparent one , R, which includes the contribution from the surface layer. Therefore, the calculated value of saturation magnetization, $M_S$, according to equation (3) would appear to be smaller. This small discrepancy between the experimental values and the ones calculated by the fanning mode is expected. The model assumes that the particle forming the chain is in single magnetic domain with "point contact" connection between adjacent particles. Also, in the calculation, n is assumed to be 12 as an approximation. Experimentally, however, higher order deviation from the model is unavoidable due to the following reasons. The magnetic structure is not exactly single-domained, because by the core-shell model, the surface layer has to be accounted for differently from the core. In addition, in the calculation of $H_C$, it is difficult to use the exact number for n. In the meanwhile, it is impossible in reality to have point contact between particles forming the chain.

## DISCUSSION

The isothermal time relaxation of the ZFC magnetization for S50 has been measured at 5 K, 50 K and 100 K with the magnetic field of 200 Oe. Using the simple equation, $M(t) = M_1 + M_2(1 - e^{-t/\tau})$, to approximate the relaxation behavior, then, $M_2/M_{ZFC}$ accounts for only 5 % at 5 K, 1.2 % at 50 K, and 2 % at 100 K, with the measuring time of two hours. Hence, the FC and ZFC data represent a reasonable approximation to the equilibrium value.

The magnetism of $Ni_3C$ phase and the effect of carbon encapsulation on pure Ni has been investigated previously [33,35]. The results in ref [34] indicate that the perfect crystalline $Ni_3C$ is expected not to have ferromagnetism. Nonetheless, because of defects in crystal or less coordination between C and Ni atoms, the $Ni_3C$ sample exhibit a very weak ferromagnetic behavior. Due to the rather weak magnetization of $Ni_3C$ and the trace amount of $Ni_3C$ existed in our sample, its magnetic effects can be neglected. The magnetic properties of the Ni phase are, therefore, not affected in the presence of little nickel carbide phase detected by the XRD in Fig. 2.

In our experiment, the saturated magnetization is seriously reduced from the bulk value. In addition, it decreases with the size of the nanochains. When comparing with some of the previously reported results mentioned in the introduction, we find that, in the cases with the reduced magnetization, the samples are not "bare". In reference [8], oleic acid (organic surfactant) is tightly bonded to the surface of $NiFe_2O_4$ powder. In reference [18], the Co nanoparticles are embedded in an amorphous matrix of $Al_2O_3$. In our experiment, the nanochain is capped with PVP. On the other hand, in the cases that the saturated magnetization retains or increases, the samples are bare. In other words, they are not in contact with other substances. Perhaps, the surface encapsulation by non-magnetic material will introduce an extra anisotropic surface potential to "trap" the surface spins. Consequently, it leads to the reduced magnetization. Although the effect of encapsulation on the magnetization is a conjecture at this stage without any firm experimental support, the effect of surface magnetic state is obvious by the present work. The saturated magnetization shows clear size dependence. It decreases with the reducing sample size.

CONCLUSION

We have performed magnetic measurements on the Ni nanochains with sizes of 50 nm, 75 nm, and 150 nm. Detailed investigations on the S50 have revealed the freezing and blocking properties attributed to the surface magnetic state and the core magnetism, respectively. These magnetic properties can be well explained by the core-shell model. In the core, the magnetic properties are similar to that of the bulk phase, except for the enhanced anisotropy due to the quasi-one-dimensional chain structure. In the surface layer, the spins exhibit spin-glass behavior, which makes enhanced contribution to the observed magnetism, e.g. enhanced saturation magnetization and coercivity, at $T < T_F$. A size-dependent effect exists on the observed magnetic properties. This can be explained within the framework of the core-shell model as well.

**Acknowledgement**

This project is financially supported by National Natural Science Foundation of

**References**

a) e-mail: cpchen@pku.edu.cn

b) e-mail: guolin@buaa.edu.cn

TABLE

|      | $M_S$ (5K) | | $M_S$ (300K) | | $M_S$ (380K) | | $H_C$ (Oe) | | |
| --- | --- | --- | --- | --- | --- | --- | --- | --- | --- |
|      | emu/g | $\mu_B$/Ni | emu/g | $\mu_B$/Ni | emu/g | $\mu_B$/Ni | 5 K | 300 K | 380 K |
| S50  | 25 | 0.26 | 11 | 0.12 | 9.5 | 0.10 | 392 | 139 | 97 |
| S75  | 44 | 0.46 | 33 | 0.35 | 30 | 0.32 | 305 | 124 | 112 |
| S150 | 52 | 0.55 | 46 | 0.48 | 42 | 0.44 | 195 | 102 | 82 |

Table 1    Saturation magnetization and coercivity for S50, S75 and S150 at 5 K, 300 K and 380 K, respectively.

Figure Captions

Fig. 1  TEM images of the nanochains. (a) Nickel chains, ~ 50 nm, capped with residual PVP. The magnification is 80 K. (b) Nickel chains, ~ 75 nm, with uniform structure. The magnification is 60 K. (c) Nickel chains with the size of 150 nm. The magnification is 20 K.

Fig. 2  XRD patterns for S50, S75 and S150. The main peaks are indexed to pure fcc nickel (JCPDS 04-0850). The remaining asterisked weak peaks, more obviously visible for S50, conform to the crystal planes of nickel carbide, $Ni_3C$, with the triagonal system (JCPDS77-0194) .

Fig. 3  FC and ZFC measurements recorded by $H_{app}$ = 90 Oe. The measurement is performed from 5 K to 395 K. For the FC measurement, the field in the cooling process is 20 kOe. The inset shows $M_{ZFC}$ in the low temperature region. The peaks around 13 K are for the freezing temperature, $T_F$. The peak position does not depend on the size, while, the peak height clearly exhibits size effect. The smaller the sample size, the larger the peak height.

Fig. 4  $M$-$T$ and $\chi$-$T$ for S50 sample. (a) ZFC data recorded by the applied field of 90 Oe, 200 Oe, 500 Oe, 800 Oe, and 1 kOe. The FC curve measured by 1 kOe is included. The FC and ZFC curves recorded at 1 kOe merge at about 50 K. (b) The susceptibility, $\chi$ = $M/H$, calculated from the data in (a). The two dashed lines are drawn to show the linear temperature behavior of $\chi$ in low magnetic field.

Fig. 5  Field dependence of the freezing temperature, $T_F$, and the blocking temperature, $T_B$ for S50. (a) External field versus freezing temperature. The dashed line represents the fitting result by the de Almeida–Thouless equation. (b) Blocking temperature versus external field. The dashed curve is for the fitting by equation (2). The anisotropic constant obtained from the fitting is, $K \sim 2.7 \times 10^6$ erg/cm$^3$.

Fig. 6  Hysteresis loops measured at 5 K, 300 K and 380 K. The inset gives the results in the saturation region.

Fig.7  Difference in the magnetization between ZFC and FC measurements, $\Delta M = M_{FC} - M_{ZFC}$, under the applied field of 90 Oe. The right Y-axis is for the coercivity.

Figure

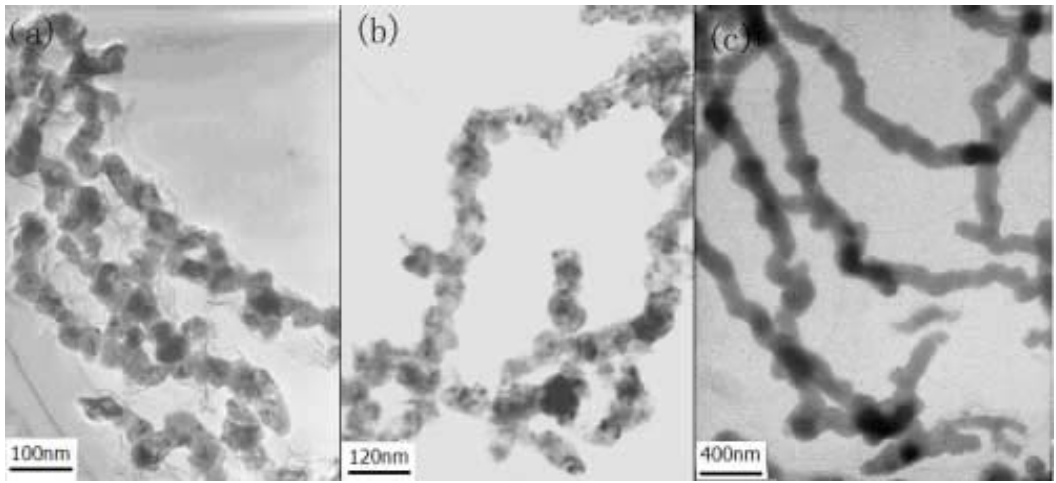

Fig 1

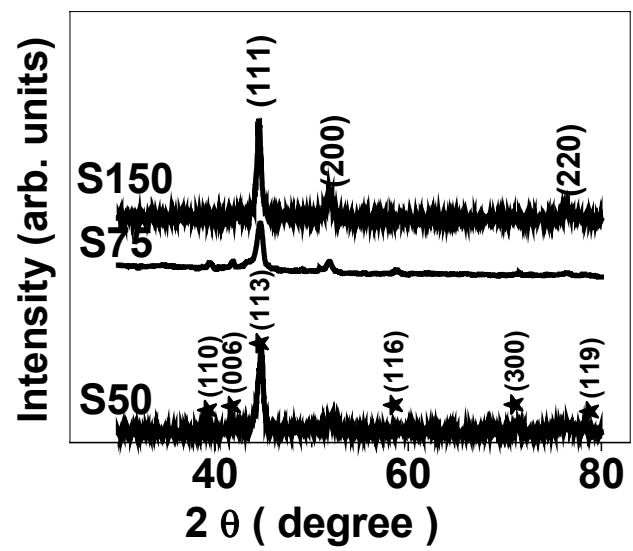

Fig. 2

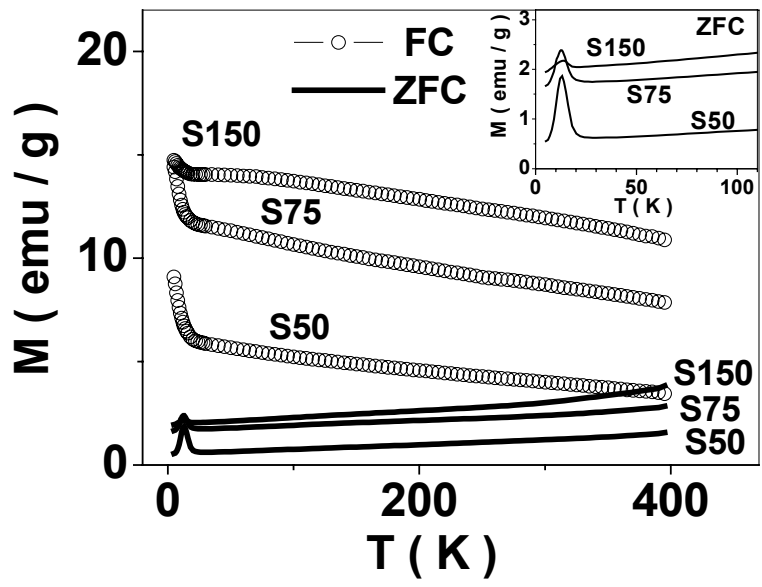

**Fig. 3**

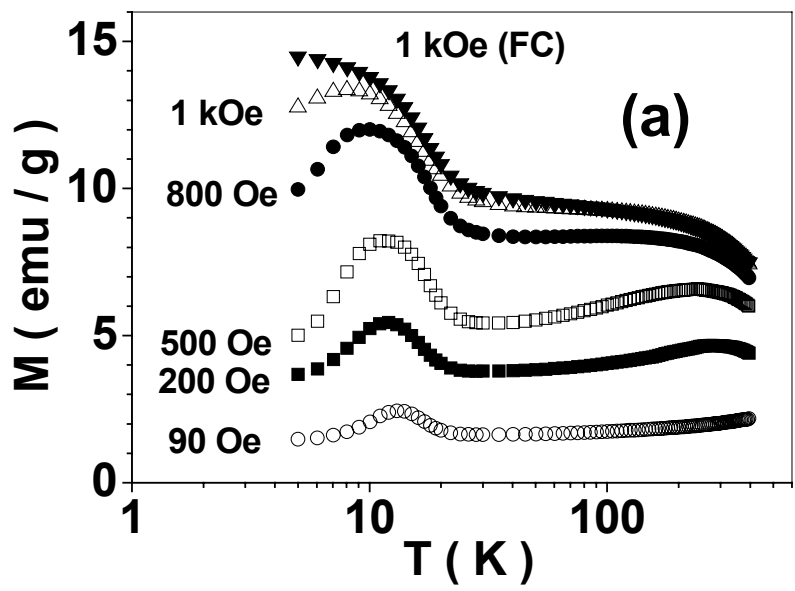

**Fig 4a**

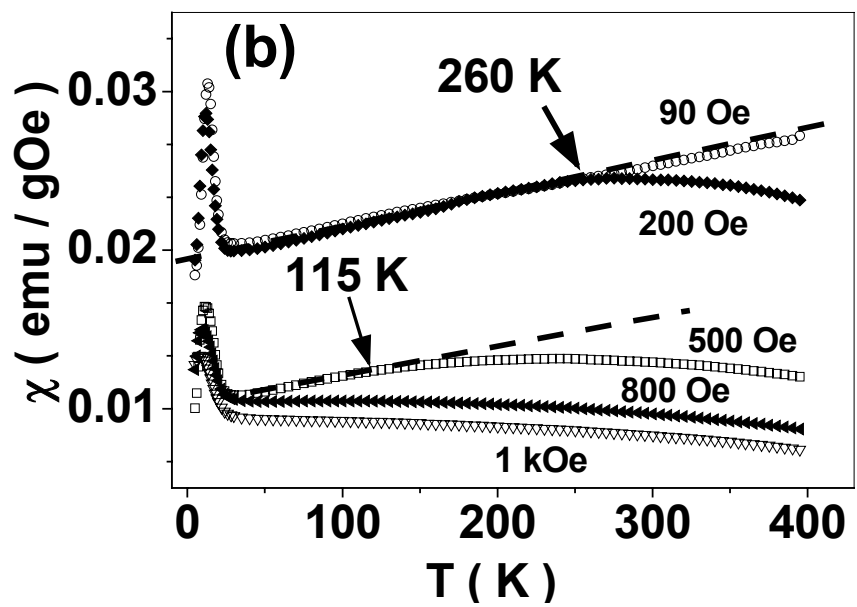

Fig. 4b

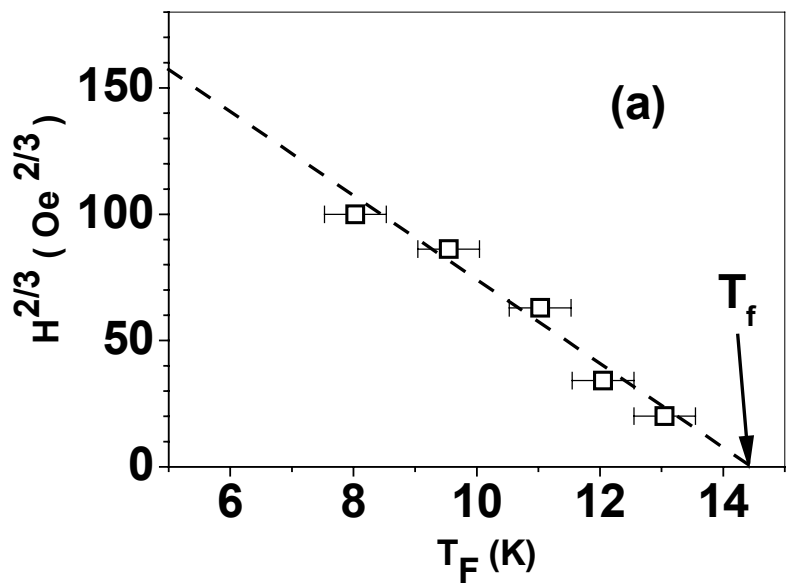

Fig. 5a

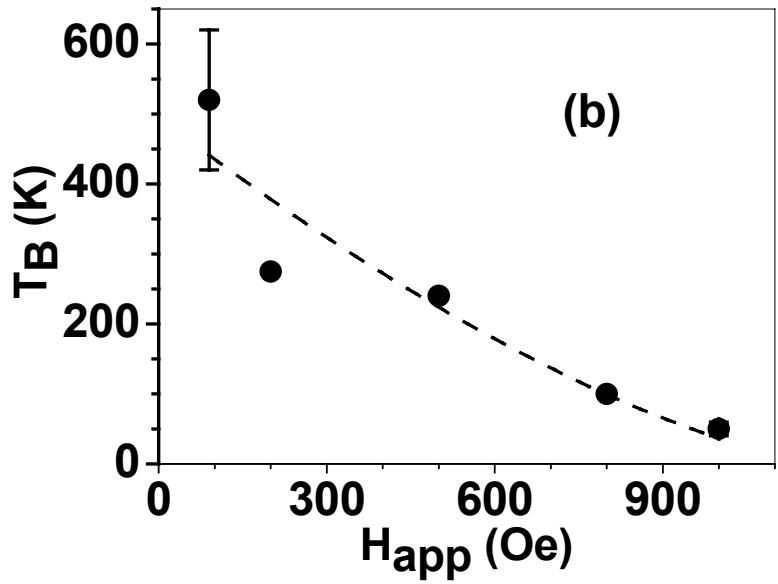

**Fig. 5b**

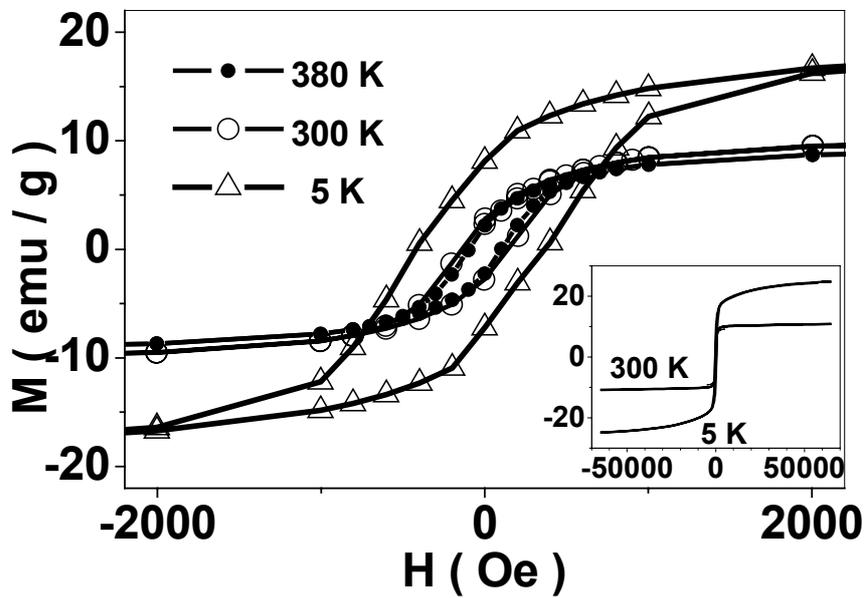

**Fig. 6**